%% file: main.tex
\documentclass[%
    reprint,
superscriptaddress,
 nofootinbib,
 amsmath,amssymb,
 aps,
 pra,
 onecolumn,
 notitlepage
]{revtex4-1} 

\usepackage[centering,hmargin=2cm,vmargin=2.5cm]{geometry}

\usepackage{graphicx}	
\graphicspath{{figures/}{./}} 
\usepackage{dcolumn}	
\usepackage{bm}			
\usepackage{hyperref}
\usepackage{ulem}
\usepackage{caption}
\usepackage{subcaption}

\usepackage{braket}
\usepackage{listings}
\usepackage{color}
\usepackage{realboxes}
\usepackage{enumitem}   
\setlist{nolistsep}

\usepackage{algorithm}
\usepackage{algorithmicx}
\usepackage{algpseudocode}
\algnewcommand\algorithmicswitch{\textbf{switch}}
\algnewcommand\algorithmiccase{\textbf{case}}
\algnewcommand\algorithmicassert{\texttt{assert}}
\algnewcommand\Assert[1]{\State \algorithmicassert(#1)}%

\newcommand{\neval}{N_{eval}}

\algdef{SE}[SWITCH]{Switch}{EndSwitch}[1]{\algorithmicswitch\ #1\ \algorithmicdo}{\algorithmicend\ \algorithmicswitch}%
\algdef{SE}[CASE]{Case}{EndCase}[1]{\algorithmiccase\ #1}{\algorithmicend\ \algorithmiccase}%
\algtext*{EndSwitch}%
\algtext*{EndCase}%

\definecolor{codegreen}{rgb}{0,0.6,0}
\definecolor{codegray}{rgb}{0.5,0.5,0.5}
\definecolor{codepurple}{rgb}{0.58,0,0.82}
\definecolor{backcolor}{rgb}{0.9,0.9,0.88}
\definecolor{mygray}{rgb}{0.8,0.8,0.8}

\def\note#1{}
\def\info#1{}

\def\eqref#1{Eq.~(\ref{#1})}
\def\figref#1{Fig.~\ref{#1}}
\def\secref#1{Section~\ref{#1}}
\def\appref#1{Appendix~\ref{#1}}

\lstdefinestyle{mystyle}{
    language=C++,
    backgroundcolor=\color{backcolor},   
    commentstyle=\color{codegreen},
    keywordstyle=\color{magenta},
    numberstyle=\tiny\color{codegray},
    stringstyle=\color{codepurple},
    basicstyle=\small,
    breakatwhitespace=false,         
    breaklines=true,                 
    captionpos=b,                    
    keepspaces=true,                 
    numbers=left,                    
    numbersep=5pt,                  
    showspaces=false,                
    showstringspaces=false,
    showtabs=false,                  
    tabsize=2,
    columns=fixed
}

\lstMakeShortInline[columns=fixed,backgroundcolor=\color{backcolor}]|

\def\snippet#1{\Colorbox{backcolor}{\lstinline|#1|}}

\begin{document}

\title{Intel Quantum Simulator:\\A cloud-ready high-performance simulator of quantum circuits}

\author{Gian Giacomo Guerreschi}
 \email{gian.giacomo.guerreschi@intel.com}
 \affiliation{Intel Labs}
\author{Justin Hogaboam}
 \affiliation{Intel Labs}
\author{Fabio Baruffa}
 \affiliation{Intel Deutschland GmbH, Feldkirchen, Germany}
\author{Nicolas P. D. Sawaya}
 \email{nicolas.sawaya@intel.com}
 \affiliation{Intel Labs}


\date{\today}


\begin{abstract}
Classical simulation of quantum computers will continue to play an essential role in the progress of quantum information science, both for numerical studies of quantum algorithms and for modeling noise and errors. Here we introduce the latest release of Intel Quantum Simulator (IQS), formerly known as qHiPSTER. The high-performance computing (HPC) capability of the software allows users to leverage the available hardware resources provided by supercomputers, as well as available public cloud computing infrastructure. To take advantage of the latter platform, together with the distributed simulation of each separate quantum state, IQS allows to subdivide the computational resources to simulate a pool of related circuits in parallel.
We highlight the technical implementation of the distributed algorithm and details about the new pool functionality. We also include some basic benchmarks (up to 42 qubits) and performance results obtained using HPC infrastructure. 
Finally, we use IQS to emulate a scenario in which many quantum devices are running in parallel to implement the quantum approximate optimization algorithm, using particle swarm optimization as the classical subroutine. The results demonstrate that the hyperparameters of this classical optimization algorithm depends on the total number of quantum circuit simulations one has the bandwidth to perform.
Intel Quantum Simulator has been released open-source with permissive licensing and is designed to simulate a large number of qubits, to emulate multiple quantum devices running in parallel, and/or to study the effects of decoherence and other hardware errors on calculation results.
\end{abstract}


\maketitle




\input{section_introduction} 
\input{section_mpi_environment} 
\input{section_scaling_experiments}
\input{section_results}   


\section{Conclusion and outlook}
\label{sec:conclusion}


We have demonstrated the functionality of Intel Quantum Simulator (IQS), a high-performance software package for simulating quantum algorithms on single work stations, supercomputers, or the cloud. Depending on the platform of choice and the problem at hand, IQS can take advantage of three operation modes: (1) all resources devoted to simulating the highest possible number of qubits, (2) processes divided into separate groups to simulate a pool of distinct circuits, or (3) using the pool of states as the stochastic ensemble needed to model noise and decoherence.


In this work we explored all three operation modes: first we launched 42-qubit simulations on the SuperMUC-NG supercomputer at LRZ and characterized the strong and weak scaling of the one-qubit gate execution. 
\note{Remove previous sentence if those results are not added}%
Then, in order to study the performance of many quantum devices operating in parallel, we used IQS to investigate the performance of particle swarm optimization (PSO) for the quantum approximate optimization algorithm (QAOA). Analyzing the results allowed us to estimate the optimal number of PSO particles for the class of problem instances studied. 
Finally, we performed a convergence study using hundreds of ensembles of stochastic circuits to describe the noise effects for systems of dimension $2^{16}\simeq 65,000$. The relatively small overhead provides a remarkable advantage over methods based on density matrix simulations, which require quadratically more memory.
We conclude by emphasizing that the two applications just discussed are particularly suitable to run on cloud platforms, where they can take full advantage of the tens of thousands of available nodes without being limited by the communication bandwidth or latency. Whether one simulates one state of many qubits or a pool of many smaller states, IQS is suitable as a standalone program or as a back-end to other quantum simulation software.



\begin{acknowledgments}
The authors thank Aastha Grover who helped in releasing Intel Quantum Simulator open source at \url{https://github.com/iqusoft/intel-qs}.
The authors acknowledge the Leibniz Supercomputing Center of the Bavarian Academy of Science (LRZ) for providing HPC resources 
and Luigi Iapichino for useful discussions about the results.
\end{acknowledgments}


\appendix
\renewcommand\thefigure{\thesection.\arabic{figure}}
\renewcommand\thetable{\thesection.\arabic{table}}
\setcounter{figure}{0}
\setcounter{table}{0}

\bigskip\noindent\makebox[\linewidth]{\resizebox{0.5\linewidth}{1pt}{$\bullet$}}\bigskip

\input{section_appendix}



\bibliographystyle{unsrt}
\bibliography{references}

\end{document}

%% file: section_introduction.tex
\section{Introduction}
\label{sec:introduction}


In the past decade there has been steady progress toward building a viable quantum computer that can be used to solve problems that classical computers cannot. Because quantum hardware is still in its infancy, the simulation of quantum algorithms on classical computers will continue to be an important and useful endeavor. This is because many technologically and scientifically relevant questions are too difficult or impractical to be answered analytically. Though many useful conclusions can be demonstrated analytically, such as the proven speedup of Shor's algorithm compared to classical factoring algorithms \cite{Shor1999} or Grover's algorithm on unstructure database search \cite{Grover1997}, most algorithmic research does benefit from numerical experiments.

The first area where numerical simulation is useful is to evaluate the performance of parameters and hyper-parameters used in quantum algorithms. For instance, the simulation of physics and chemistry problems involved many choices regarding how to encode the problem to a set of qubits \cite{Sawaya2019a}, for which it is usually not obvious which approach will be most efficient without performing numerics. Further, most variational algorithms --- whether the variational quantum eigensolver for finding Hamiltonian eigenvalues \cite{Peruzzo2014} or variants of the quantum approximate optimization algorithm for solving combinatorial problems \cite{Farhi14_qaoa_orig} --- involve a classical heuristic optimization routine that, for all intents and purposes, must be analyzed numerically.

The other primary reason for numerical simulation of quantum algorithms is to study the effects of errors. Despite enormous progress in reducing the effects of environmental noise \note{refs}and in perfecting the fidelities of gate operations, it appears certain that all near-term quantum devices will exhibit errors that cannot be corrected without sophisticated error correction schemes such as the surface code \cite{Kitaev2003}. This highlights the need for numerical simulations of quantum algorithms running on error-prone hardware \cite{Sagastizabal2019,Guerreschi2019}. Such simulations not only help draw conclusions about the robustness of particular algorithmic choices, but can also guide hardware design and gate compilation \cite{Tannu2019,Lao2019a}, since different choices may lead to qualitatively different errors.

Quantum circuits are hard to simulate classically since the computational cost scales exponentially with the number of qubits.
\note{Several implementations [https://www.quantiki.org/wiki/list-qc-simulators] targeting single node shared memory systems are present and very few are targeting Supercomputing systems [https://www.nature.com/articles/s41598-019-47174-9].}%
Notably, though there are classes of algorithms that scale more favorably for some set of quantum circuits --- such as tensor network \cite{Markov2008_tn,Fried2018_qtorch,McCaskey2016_tnqvm,gray2018_quimb,Villalonga2019} or path integral \note{ref} methods --- these methods still scale exponentially in the general case. Several high-performance quantum circuit simulators have been reported, including full state vector codes built for CPUs \cite{Niwa2002, DeRaedt2007, Smelyanskiy2016, Haner2017, Khammassi2017, LaRose2018, Jones2019, DeRaedt2019_48qub} and/or graphics processing units \cite{Gutierrez2010, Amariutei2011, Zhang2015_gpu, Haner2017, Savran2018_shor_gpu}, and those that use a mix of algorithm types \cite{Pednault2017, Chen2018_64qubit}.

In this manuscript we present a new version of Intel Quantum Simulator (IQS) and use it to emulate a hybrid quantum-classical algorithm. Also known as qHiPSTER or the Quantum High-Performance Software Testing Environment, IQS is a massively parallel simulator of quantum algorithms expressed in the form of quantum circuits. Its original version was coded for High-Performance Computing environments, with the goal of allowing large scale simulations. In 2016 it was used to simulate the full state of 40 \cite{Smelyanskiy2016} and later 42 qubits \cite{Boixo2018}.

The second version of Intel Quantum Simulator has just been released, open source, at the address \url{https://github.com/iqusoft/intel-qs}. While preserving the HPC core of the original implementation, it includes several new features that extend its application to cloud computing environments. In particular, Intel Quantum Simulator can divide the allocated computational resources into \textit{groups}, each dedicated to the simulation of a distinct quantum circuit. The communication between these group of processes is minimal, and each separate group uses the distributed implementation from the original release to store and manipulate its quantum state. We expect that at least two use cases will profit massively from this extension. First, when multiple circuits or variants of the same circuit have to be run in parallel (think for example of variational algorithms in conjunction with classical optimizers like the genetic algorithm or particle swarm optimizers). And second, when stochastic methods are used to include noise and decoherence in the simulation.
\note{rephrase previous sentence?}%
One can easily envision situations as those just mentioned in which a pool of hundreds or thousands of states is required to accelerate simulation. Cloud computing platforms, with (tens of) thousands of nodes, are an ideal choice to run these workloads.

In addition to the new features just discussed, the new release includes robust unit testing to verify the proper installation and functioning of IQS and, for developers, to test the compatibility of novel features with the released code.
Finally, we focused on lowering the user's learning barrier with an automatic installation process and extended tutorials, and improving the simplicity of use by providing Python integration and a Docker container option. Our goal in releasing this version of the software is that IQS may be used as a standalone program or as a backend to other quantum computing frameworks like Xanadu's Pennylane \cite{Pennylane}, IBM's Qiskit \cite{Qiskit}, Rigetti's  Forest \cite{Forest}, Google's Cirq \cite{Cirq}, Microsoft's Azure Quantum \cite{Azure}, ProjectQ \cite{Projectq}, Zapata's Orquestra \cite{Orquestra}, Amazon's Braket \cite{Braket}, and others.

In this article, we begin by describing the basic usage of IQS and its software structure. In \secref{sec:scaling_experiments}, we present benchmarks for large scale simulations of up to 42 qubits. \secref{sec:results} describe two situations that take advantage of simulating a pool of circuits: the emulation of a variational protocol that uses many quantum processors in parallel and the simulation of circuits exposed to noise. Finally we draw some conclusions and provide an outlook.


%% file: section_mpi_environment.tex
\section{Software description}
\label{sec:software}


Intel Quantum Simulator, both in its initial version and latest release, takes advantage of the full resources of an HPC system, due to the shared and distributed memory implementation.
The first situation is when several processors, or a processor with multiple computing cores, have access to the same memory and the operations need to be performed without a specific sequential order. This opportunity for parallelism is best exploited with OpenMP.
The second opportunity for parallelism arises when a relatively small amount of memory requires a lot of computation or, as in the case of storing quantum states, a large amount of memory cannot fit in a single machine or node. In this case, one needs to explicitly consider the communication pattern between the different processes and adopting the Message Passing Interface is a necessity.

\begin{figure}
  \centering
    \includegraphics[width=0.9\textwidth]{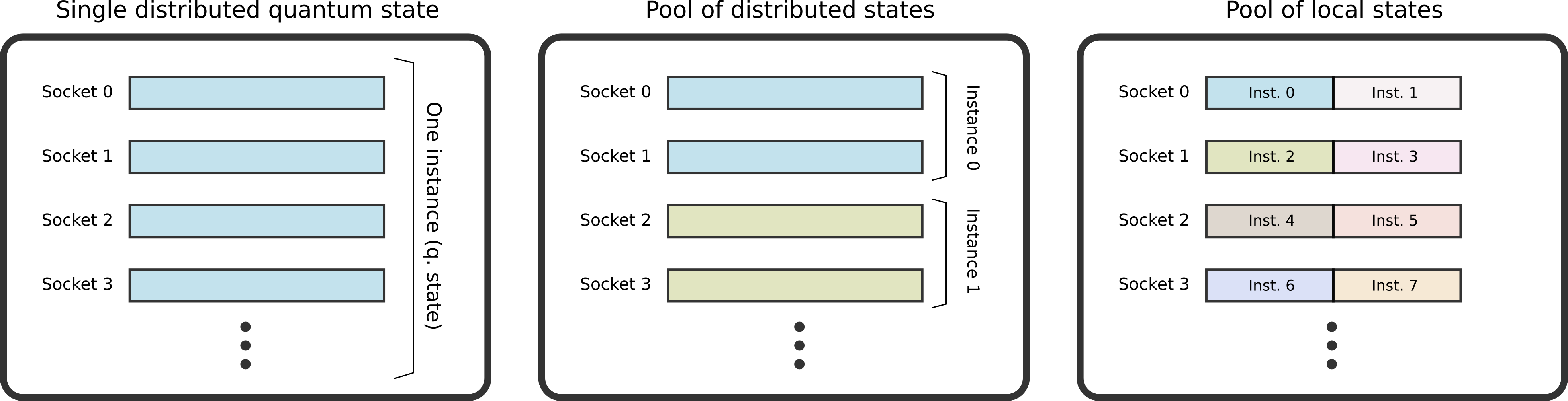}
  \caption{Depending on the nature of the research question being considered, there are several ways to run IQS. If one needs to simulate a single instance of a circuit for the highest possible qubit count, then one would distribute the quantum state across all available computing nodes or sockets \textbf{(left panel)}. If one needs to consider a pool of states for either simulating many circuits in parallel or describe noise effects via stochastic simulations (or possibly both situations at the same time), then one may simulate multiple distributed \textbf{(center panel)} or local \textbf{(right panel)} quantum states in parallel.}
  \label{fig:mpi_modes}
\end{figure}

In the new release of IQS we have set the MPI environment for allowing multiple quantum circuits to be simulated in parallel. We divide the computing processes into groups, each dedicated to storing a single quantum state and update it according to the action of a specific circuit. Each new state can still profit from the shared and the distributed implementation of the code (MPI+OpenMP), which has been previously implemented in the original version of the simulator. The use cases are illustrated in \figref{fig:mpi_modes} where the graphics clarifies that a single state can be stored using all nodes, a subset of them, or even part of a single computing node. In this section we first introduce the basic methods that allow IQS to initialize, evolve and extract information from quantum states, then we discuss the distributed implementation of a single state and finally explain the parallel simulation of multiple circuits.

\subsection*{Scope and basic use}

IQS is designed to simulate the dynamics of multi-qubit states in quantum circuits. The state is assumed to be pure and its unitary dynamics a consequence of the application of one- and two-qubit operations. IQS provides methods to extract information from the state at any intermediate point or at the end of quantum circuits. There are three fundamental parts in the simulation of quantum circuits: state initialization, evolution, and measurement. We illustrate the basic methods of IQS with short snippets of code which are part of a longer C++ program contained in the release among the examples. The same syntax applies to the Python interface of IQS%
\footnote{The Python interface of IQS is currently limited to single-process execution (i.e. no MPI).}.

The central object of IQS is called \snippet{QubitRegister} and can be thought as the quantum state of the qubits composing the system of interest. In its declaration, we need to specify the number of qubits in order to allocate a sufficient amount of memory to describe their state. Then we can initialize the state to any computational basis state, uniquely identified by its index. Other methods allow for initializing the state randomly or, for example, as the balanced superposition of all computational states.

\begin{lstlisting}
  int num_qubits = 4;
  QubitRegister<std::complex<double>> psi(num_qubits);
  std::size_t index = 0;
  psi.Initialize("base", index);
\end{lstlisting}

The dynamics is generated by applying one- and two-qubit gates. IQS gives the option of defining custom gates, but also provides a large choice of the most common gates like single-qubit rotations or the Hadamard gate. The two qubit gates are in the form of controlled one-qubit gates, meaning that the desired operation is applied to the target qubit conditionally on the control qubit being in $\ket{1}$. The special case of the conditional Pauli X, also called CNOT gate, clarifies why there is no need of arbitrary two- or multi-qubit gates: any unitary evolution can be approximated to arbitrary precision by a sequence of one-qubit and CNOT gates. IQS is suitable for implementing multi-qubit operations%
\footnote{For example, the latest IQS release include methods specialized to the emulation of circuits for the quantum approximate optimization algorithm. These circuits are reduced to one-qubit gates and a single global operation per step.}, but the definition of custom multi-qubit gates requires a very good understanding of its internal implementation (see next subsections).

\begin{lstlisting}[firstnumber=5]
  // One qubit gates: Pauli X on qubit 1 and Hadamard on qubit 0
  psi.ApplyPauliX(1);
  psi.ApplyHadamard(0);
  // Two-qubit gate of conditional form: apply Pauli X on qubit 0 conditioned on qubit 1
  int control_qubit = 1;
  int target_qubit = 0;
  psi.ApplyCPauliX(control_qubit, target_qubit);
\end{lstlisting}

Finally the qubits can be measured in the computational basis one at a time or in larger groups. While in actual realization of the quantum experiments a measurement returns only one of the possible outcomes according to a state-dependent probability distribution, with simulators one can compute the full statistics of outcomes without the need of re-running the experiment. For example, IQS provides methods to compute the probability of finding a certain qubit in state $\ket{1}$ or evaluate the expectation value of multi-qubit observables like products of Pauli matrices (not shown below).

\begin{lstlisting}[firstnumber=15]
  // Compute the probability of observing qubit 1 in state |1>
  int measured_qubit = 1;
  double probability = psi.GetProbability(measured_qubit);
  // The expectation value of <Z1> can be computed from the above probability
  double expectation = -1 * probability + 1 * (1-probability);
\end{lstlisting}

\subsection*{Distributed implementation}

Here we describe how the quantum state is defined and stored inside the IQS objects. The current algorithm is based of the original implementation of Intel Quantum Simulator~\cite{Smelyanskiy2016}, so here we summarize it in order to provide  a self-contained description of our simulator.

To fully describe an arbitrary state of $n$ qubits, one needs to store $2^n$ complex numbers corresponding to the probability amplitudes with respect to the computational basis. For $n$ as low as 30, simply storing all the amplitudes fills up $2^{3+1+30} \text{Byte} \simeq 17$~GB of memory ($2^3=8$ bytes per double-precision number and a factor $2$ since the probability amplitudes are complex).
To enable the fast simulation of circuits involving more than 30 qubits, one needs to divide the state between multiple processes, each with its own dedicated memory. IQS assumes that $P=2^p$ processes are used, each storing $2^{n-p}$ amplitudes and satisfying $p<n$. If $P$ is not a power of two, IQS considers an effective number of nodes equal to $2^p$ with $p=\lfloor\log_2(P)\rfloor$.

As we explain in the next paragraphs, all operations involving only the first $m=n-p$ qubits do not require communication between processes, while MPI communication is needed when performing operations on the last $p=n-m$ qubits. Therefore we refer to the qubits with index $0\leq q<m$ as ``local'' and those with index $m\leq q<n$ as ``global''. However it is important to realize that even the partial state of a local qubit can be fully known only by accessing all $2^n$ amplitudes distributed among all $2^p$ processes.

It is informative to analyze how one-qubit gates are implemented in IQS. Any quantum state can be written as a vector with complex entries $\{ \alpha_i \}_{i=0,1,\dots,2^n-1}$, so it is convenient to express the index $i$ in binary notation as $i_{n-1} \dots i_2 i_1 i_0$ with $i_q\in\{0,1\}$ and such that $i=\sum_{q=0}^{n-1} i_q 2^q$. In this way it is straightforward to obtain both the process number $p(i)$ and index of the local memory $\ell(i)$ corresponding to the $i$-th amplitude $\alpha_i$:
\begin{equation}
\label{eq:process_index}
     p(i) = \sum_{q=m}^{n-1} i_q \, 2^{q-m} \ ,
     \quad\quad
  \ell(i) = \sum_{q=0}^{m-1} i_q \, 2^{q}
\end{equation}

\begin{figure}[t]
\centering
\includegraphics[width=0.7\textwidth]{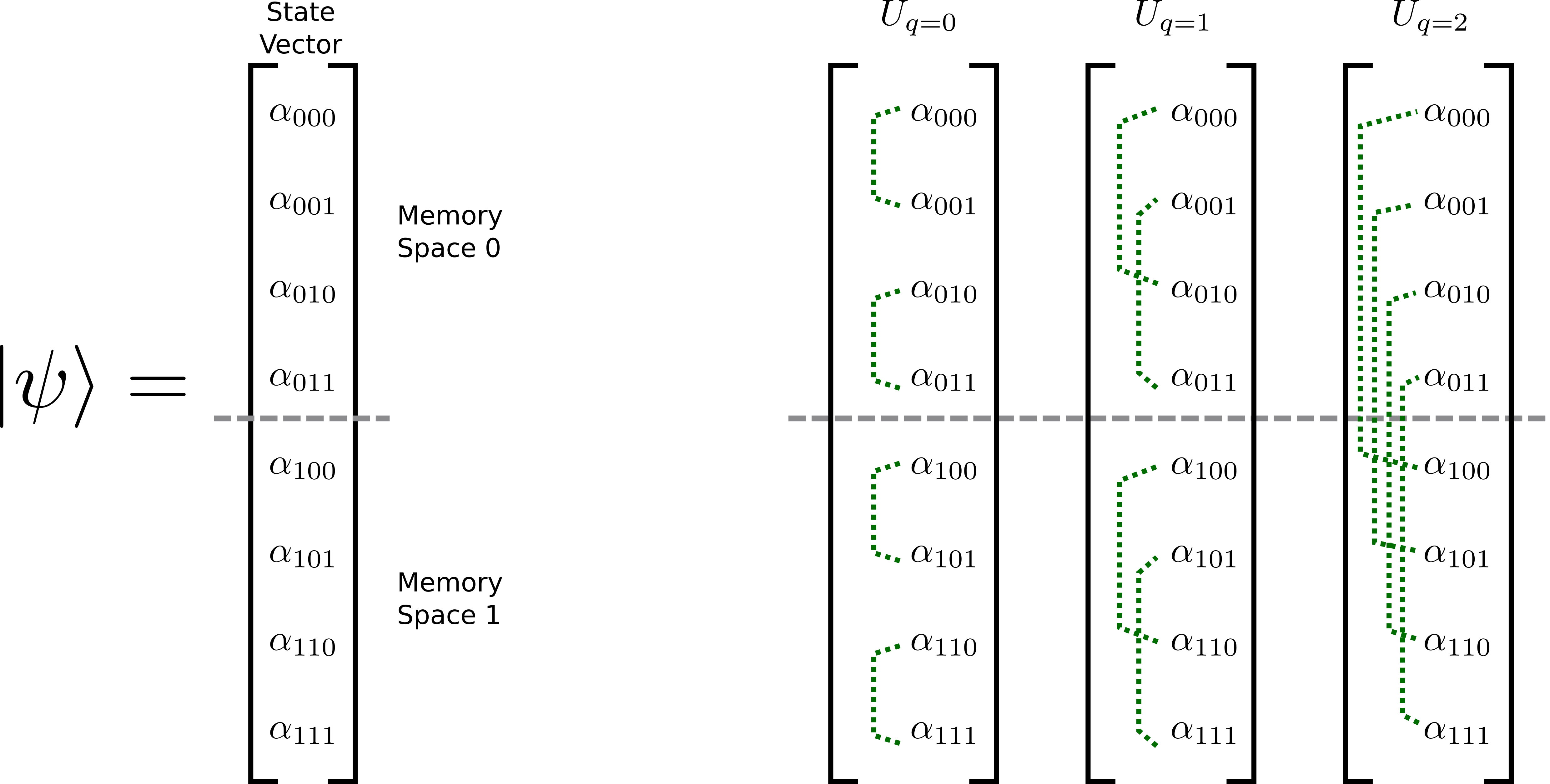}
\caption{\textbf{Left:} Quantum state of three qubits stored as a vector of $2^3=8$ complex amplitudes $\alpha_{i_2 i_1 i_0}$:
$\ket{\psi}=[\alpha_{000}, \alpha_{001}, \alpha_{010}, \dots, \alpha_{111}]^\text{T}$.
The state is distributed over 2 processes, each storing half of the amplitudes.
\textbf{Right:} Illustration of the computation scheme to simulate one-qubit gates. Observe the qualitative change depending on the qubit involved in the operation: At a critical qubit number, communication between memory spaces (\textit{i.e.} processes or MPI ranks) is required. In this 3-qubit case, communication between memory spaces is required for $q=2$ but not for $q=0,1$.
}
\label{fig:3qub_ex}
\end{figure}

Consider the one-qubit gate acting on qubit $q$ and defined by the $2\times 2$ unitary matrix $U$:
\begin{equation}
  U = \begin{pmatrix}
U_{00} & U_{01}\\
U_{10} & U_{11}
\end{pmatrix} \, .
\end{equation}
Its action on the quantum state can be written as
\begin{align}
  \alpha^\prime_{\star \dots \star 0_q \star \dots \star} &=
      U_{00} \, \alpha_{\star \dots \star 0_q \star \dots \star} +
      U_{01} \, \alpha_{\star \dots \star 1_q \star \dots \star} \nonumber \\
  \alpha^\prime_{\star \dots \star 1_q \star \dots \star} &=
      U_{10} \, \alpha_{\star \dots \star 0_q \star \dots \star} +
      U_{11} \, \alpha_{\star \dots \star 1_q \star \dots \star}
\end{align}
where $\star$ refer to any bit value. The expression above means that the entries are updated in pairs independently of the values of the other amplitudes. From eq.~\eqref{eq:process_index} it is clear that when $q<m$ the connected pairs of entries are stored in the memory of the same process and can be updated without inter-process communication, as illustrated in \figref{fig:3qub_ex}.

\begin{figure}[h]
\centering
    \includegraphics[width=0.9\textwidth]{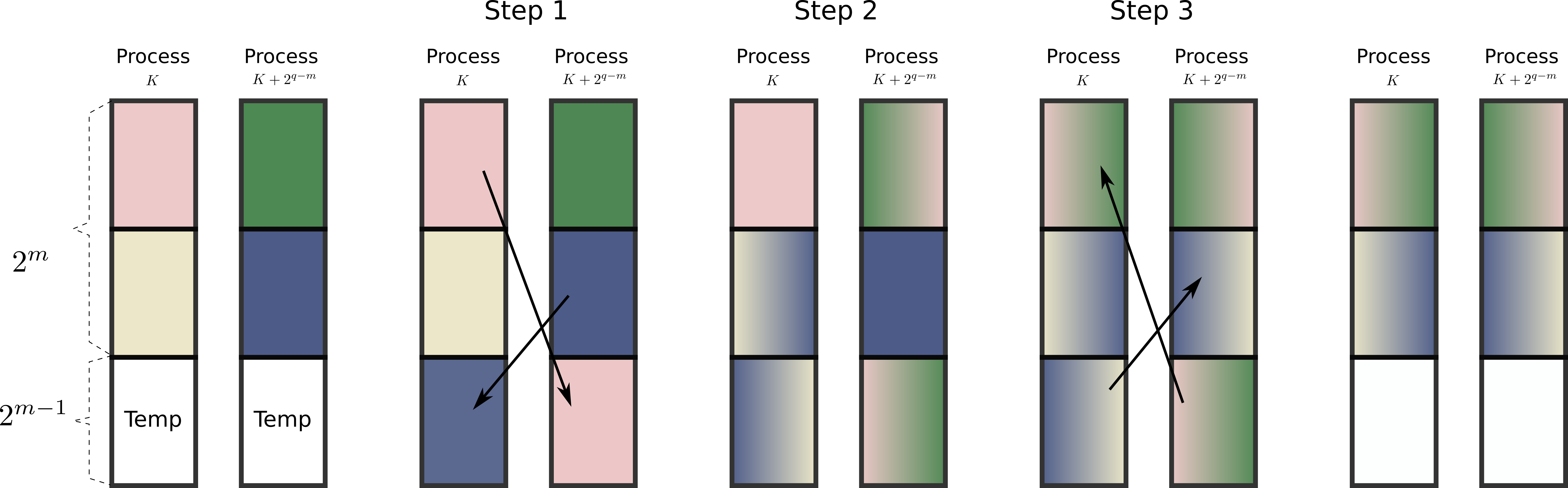}
\caption{Communication scheme for implementing a one-qubit gate on qubit $q$, with $q>m$. 
Time flows from left to right. \textbf{Step 1:} Each MPI task sends half of its local memory to its communication partner, identified by an index difference of $2^{q-m}$. \textbf{Step 2:} The computation is equally split between the two processes and involves only local memory. \textbf{Step 3:} At the end the updated information is sent back and the state is updated. This follows the original IQS implementation  \cite{Smelyanskiy2016} that was first described in \cite{Trieu2009}.}
\label{fig:MPI_communication}
\end{figure}

The situation differs when $q\geq m$. In this case the two entries belong to the memory of two distinct processes, specifically to those with index $p(i)$ and $p(i+2^q)=p(i)+2^{q-m}$ respectively (here $i$ is such that $i_q=0$). Inter-process communication is therefore required and we adopt the same scheme as in the original IQS implementation \cite{Smelyanskiy2016, Trieu2009}. It is briefly summarized in \figref{fig:MPI_communication} and its caption.

In addition to one-qubit gates, IQS implements distributed two-qubit gates of the controlled form, meaning that the one-qubit gate $U$ is applied to target qubit $t$ conditionally on the control qubit $c$ being in state $\ket{1}$. The communication pattern depends on the control and target qubits being local or global and if $t>c$ or not.

\subsection*{Pool of multiple states}

The most consequential change in the IQS implementation compared to its original release is the ability of dividing the processes into groups using the MPI function \snippet{MPI_Comm_create_group}. Each group can be used to store a quantum state, possibly in a distributed way if the group itself is composed by more than one process. Now, when a \snippet{QubitRegister} object is created, it actually initializes a state in each group of processes: we call ``pool'' the collection of such states. In addition, when a method of the form \snippet{ApplyGate} is called, the gate is actually applied to each and every state of the pool.

We clarify this concept with a concrete example. Consider that we have 80 processes and want to work with 10 quantum states. One option is to group all processors together and declare 10 \snippet{QubitRegister} objects: here, each state is distributed over $64=2^6$ processes (since 80 is not a power of 2) and the circuit's gates must be specified for each of the 10 states separately. The second option is to divide the processes in 10 groups of 8 processes each and create a single \snippet{QubitRegister} object: here, each state is distributed over $80/10=8=2^3$ processes and each gate is by default applied to every state.

There are two important observations: the first one is that defining a non-trivial pool of states may take advantage of the available processes in a more effective way. The second is that each state of the pool is naturally subjected to the same quantum circuit. The latter characteristic, if strictly enforced, would make the simulations redundant: we would simulate over and over the same identical evolution. However it is possible to differentiate the applied circuit for each of the state in the pool and the relevant commands are discussed in \appref{app:sec:multi-state}. Moreover, there are cases in which simulating closely related circuits is required and what seemed a limitation actually becomes a beneficial feature.

Here we discuss two of these situations and present the corresponding results in \secref{sec:results}. Code snippets are discussed in \appref{app:sec:multi-state}.
\begin{itemize}
  \item In Variational Quantum Algorithms (VQA) a quantum circuit composed of parametric gates is optimized to prepare states with desired properties, often related to having large overlap with the ground state of certain observables. During the optimization, the same circuit is simulated over and over with the only difference being the value of its parameters (think of them as the angle of one-qubit rotations). Within the pool functionality, it is easy to assign different parameter values to the circuit simulated by the distinct states in the pool. This approach greatly speedup the overall simulations of VQA protocols based on several classes of optimizers, like genetic algorithm, swarm particle optimization, or gradient-based methods.
  \item IQS is a simulator of unitary dynamics in which each state is pure. Nonetheless it is possible to use IQS to simulate the effect of noise and decoherence during the circuit by means of introducing stochastic perturbations to the ideal circuit and averaging over the ensemble of ``perturbed'' circuits \cite{Smelyanskiy2016}. Formally, this approach is based on the unraveling of master equations into stochastic Schr\"odinger equations in the circuit-model formalism \cite{Bassi2008} and corresponds to the introduction of additional ``noise gates'' in the form of one-qubit rotations with stochastic rotation angles. IQS provides specialized methods to apply these noise gates that automatically varies their rotation angles over the pool's states.
\end{itemize}

\info{Resource on groups and Communicators:\\
\url{https://mpitutorial.com/tutorials/introduction-to-groups-and-communicators/}}


%% file: section_scaling_experiments.tex
\section{Scaling Experiments}
\label{sec:scaling_experiments}

In Ref.~\cite{Smelyanskiy2016}, Smelyanskiy and coauthors analyzed the simulator performance on the distributed system Stampede provided by the Texas Advanced Computing Center (TACC). They demonstrated the weak and the strong scalability of the code up to 40 qubits using $1024$ compute nodes. They also performed single and multi-node performance measurements of one- and two-qubit gates, and  
of a complete quantum circuit, namely the Quantum Fourier Transform. Since the core implementation of IQS did not change for the latest release, we consider those results still valid.

In the current release of IQS, we provided sample codes to run the simulation of one-qubit operations by varying the total number of qubits in the state or the index of the qubit involved in the gate. This allows the users to benchmark IQS execution times on HPC systems with the scope of analyzing the strong and the weak scaling of the simulator. We have used such sample scripts to run the following experiments launched on the SuperMUC-NG\footnote{https://doku.lrz.de/display/PUBLIC/Hardware+of+SuperMUC-NG}
HPC system hosted by the \textit{Leibniz Supercomputing Center of the Bavarian Academy of Science} (LRZ). SuperMUC-NG consists of $6,480$ compute nodes, each equipped with 2 socket Intel\textsuperscript{\textregistered} Xeon\textsuperscript{\textregistered} Scalable Processor 8174 CPU. The total amount of CPU cores is $311,040$ and the total distributed memory is $719$ terabytes. Each single node is a two sockets system of $24$ cores each with $96$GB of shared memory. In the next sections, we present strong and weak scaling results up to $2048$ nodes, which corresponds to $98,304$ CPU cores and a total memory of $196$TB. 

\subsection*{Strong scaling}
\begin{figure}[h]
\centering
\includegraphics[width=0.4\textwidth]{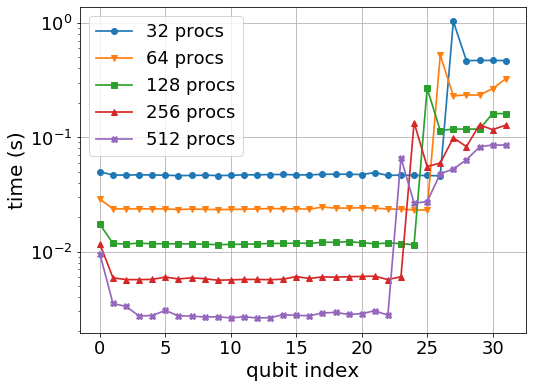}
\hspace{4mm}
\includegraphics[width=0.4\textwidth]{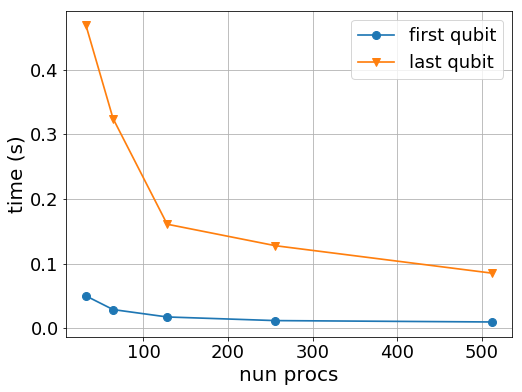}
\caption{Strong scaling of 32-qubit simulations using from 32 to 512 MPI processes. Each MPI process is running on one socket of SuperMUC-NG, which means 2 MPI tasks per node. Each socket is fully populated by 24 OpenMP threads.
\textbf{Left panel:}
Time to execute a random one-qubit gate as a function of the qubit involved. Different colors correspond to different number of MPI processes. When the gate is executed on qubit $q=n-p$ (with $2^p$ being the number of processes and $n=32$), the communication between the MPI tasks is happening within the nodes (intra-node) and not between the nodes (inter-node). For the later qubits the communication is mostly inter-node. 
This reflects the peak behavior we observe in our measurement, which has been confirmed also by our MPI ping-pong test benchmark for messages of that size.
\textbf{Right panel:}
Time to execute a random one-qubit gate on the first ($q=0$) or last ($q=31$) qubit. The computational time difference is due to the additional communication required for the last qubit to be updated.
}
\label{fig:strong_scaling}
\end{figure}

In the strong scaling analysis, we fixed the problem size (the number of qubits to run) and scaled up the computational resources. The total speedup is limited by the fraction of the serial part of the code that is not amenable to parallelization. In \figref{fig:strong_scaling} we show the speedup of single-qubit operations for simulations of 32-qubit system. We have implemented one-qubit gates defined by a random $2\times 2$ matrix. The gate is then applied to all the qubits involved in the simulation. 

\begin{figure}[t!]
\centering
\includegraphics[scale=0.4]{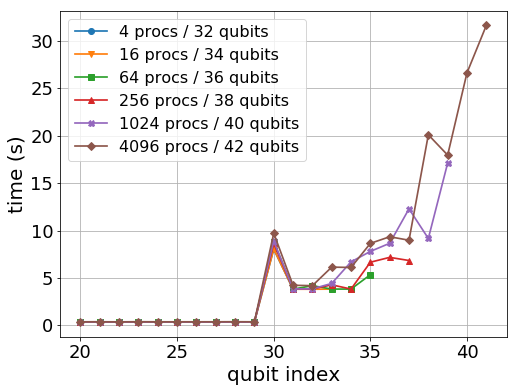}
\caption{Weak scaling study of the time needed to apply one-qubit gates depending on the involved qubit. Different colors correspond to different number of MPI processes and number of qubits. For any additional qubit in the simulation, we increase the computational resources by a factor of 2 since the memory required by the state representation also increases by a factor of 2. For small qubit indices, we observe exactly the same computational time using larger numbers of qubits and resources. For larger qubit indices, the difference is mostly due to the communication overhead. The peak we observe at qubit index 30 is due to the intra-node communication between the
2 sockets of one node.}
\label{fig:weak_scaling}
\end{figure}

\subsection*{Weak scaling}
In \secref{sec:software} we described the internal representation of quantum states used by IQS and highlighted the fact that simulating an extra qubit implies doubling the allocated memory. This consideration can be included in the numerical analysis of the so-called weak scaling. The idea is that, while the number of processes increase, the simulation size also increases and in such a way that the memory amount and computing effort per process stays (ideally) constant.

We launched simulations of systems from 32 to 42 qubits using from 4 to 4096 processes. The largest job used 2048 nodes of the SuperMUC-NG system. The expectation of a scale-invariant behavior is confirmed by our study and presented in \figref{fig:weak_scaling}. 

%% file: section_results.tex
\section{Particle Swarm / QAOA Simulation}
\label{sec:results}

Having discussed the functionality and implementation of IQS, we now consider an illustrative application. As quantum hardware improves, it will be possible to run many small-scale quantum computers at the same time. Though there may not be entanglement across the devices, one may run the devices in a parallel fashion in order to more quickly solve variational quantum problems. Each quantum device would be calculating an objective function for a different set of parameters, with a classical optimization step using the results of all the devices. The behavior of such an algorithm may be analyzed numerically using IQS.

Besides studying the behavior of a specific algorithm, a purpose of this section is to demonstrate the IQS `pool of states' functionality. Because the classical optimization loop we use has low overhead, the computational speedup up of this distributed IQS simulation---compared to a single-process simulation---will be approximately equal to the number of processes used in the simulation.

\subsection*{QAOA with swarm particle optimization}

We used IQS to perform this task of simulating many quantum computers running in parallel. The simulation demonstrates one example of the variety of simulation types that may be performed with the software. The variational quantum algorithm we chose is the quantum approximate optimization algorithm (QAOA) \cite{Farhi14_qaoa_orig} for the Max-Cut problem on 3-regular graphs, an extensively studied problem in the quantum algorithms community \cite{Farhi14_qaoa_orig, Farhi14_qaoaboundoccur, Farhi16_qsuprqaoa, Wecker16_training, Lin16_qaoa, Guerreschi17_qaoa_opt, Guerreschi2019}. For the classical optimization procedure we use the particle swarm optimization (PSO) algorithm \cite{Kennedy1995,Pedersen2010}, where we implement each `particle' as one virtual quantum device.

\begin{figure}[t!]
\centering
    \includegraphics[width=0.5\textwidth]{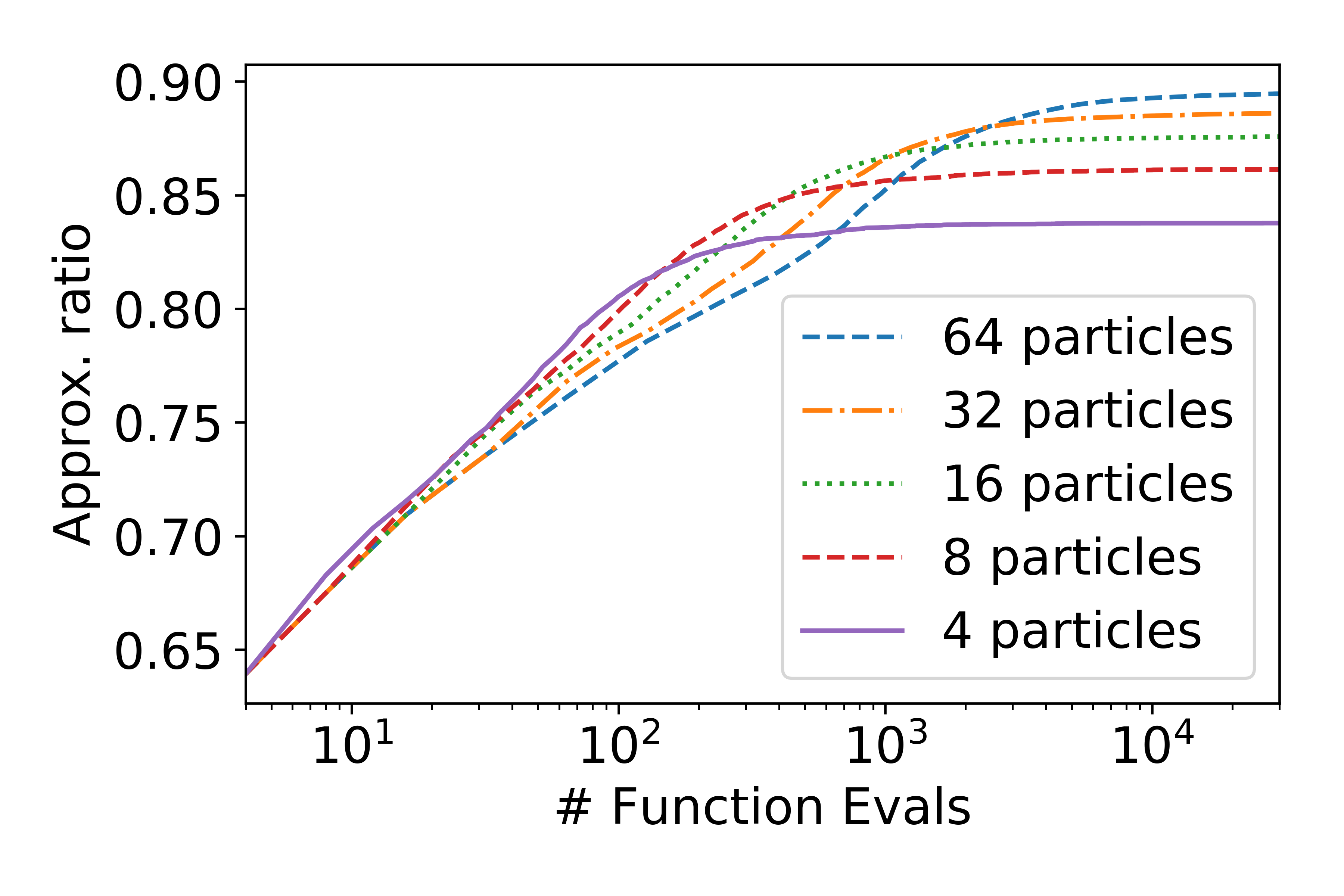}
\caption{ Approximation ratio versus the number of total function evaluations, while running PSO for QAOA/Max-Cut on 3-regular graphs of 18 vertices (18 qubits) for varying particle counts. Results are averaged from 300 different graphs and random initial conditions. The horizontal axis gives the number of function evaluations, \textit{i.e.} the number of quantum circuit emulations and $\langle H_{MaxCut}\rangle$ evaluations run on IQS. The vertical axis gives the approximation ration, equal to $ \bra{0} U^\dag_{circ} H_{MaxCut} U_{circ} \ket{0}$ divided by the exact MaxCut solution. The higher the particle count, the larger the number of function evaluations per PSO time step. Note that the appropriate choice of particle count depends on how many function evaluations one wishes to perform. Standard deviations (not shown in figure) calculated over the set of graphs are large compared to typical difference between means, often higher than 0.05.}
\label{fig:swarm-dynamic}
\end{figure}

The particular PSO implementation we used was taken from reference \cite{Pedersen2010}. For each particle we first set random initial positions drawn uniformly from $[0,2\pi)$, where these positions are each a set of parameter vectors $\{\vec \theta_0, \vec \theta_1, \cdots, \vec \theta_{R-1} \}$ for $R$ particles. 
Each unique position $\vec \theta_k$ produces a unique output from the objective function $\mathcal L(\vec \theta)$ $=\langle H_{MaxCut}\rangle$, where $H_{MaxCut}$ is the Max-Cut Hamiltonian. Each particle is given an initial velocity $\vec v_k$ also drawn uniformly from $[0,2\pi)$. The particles are propagated for one time step based on their velocities, after the velocities have been updated with the formula
%
\begin{equation}
\vec v_{k} \leftarrow \omega \vec v_k + \phi_p r_p[\vec \theta_k - \vec \theta_{k,(best)}] + \phi_g r_g [\vec \theta_k - \vec \theta_{(global)} ]
\end{equation}
where $\omega$, $\phi_p$, and $\phi_g$ are arbitrary constants; $r_p$ and $r_g$ are random numbers drawn each step from uniform distribution [0,1]; $\vec \theta_{k,(best)}$ is the best position that particle $k$ has discovered thus far; and $\vec \theta_{(global)}$ is the best position discovered by the entire swarm. For this work, we set the parameters to $\omega=0.66$, $\phi_p=1.6$, and $\phi_g=0.62$, as there is numerical evidence that these parameters perform well for some optimization landscapes and hyperparameters \cite{Pedersen2010}. The algorithm is naturally parallelizable, as the only cooperation between quantum devices involves broadcasting each device's $\mathcal L$ in order to modify the velocities. The formula shows that after each time step, each new velocity $\vec v_k$ is determined by three terms: a damping term ($\omega$), an acceleration term based on the best previous position of the $k$th particle ($\phi_p$), and a second acceleration term based on the best position found so far by the entire swarm ($\phi_g$).


In the procedure described above, we make the implicit assumption that systematic error does not differ greatly between devices. One feature of variational quantum algorithms is that they are robust to any systematic constant errors inherent to a given device. This is because $\max \mathcal L (\vec \theta) $ $= \max \mathcal L (\vec \theta + \vec \varepsilon)$, \textit{i.e.} the maximum of the objective function is independent of any systematic error $\vec \varepsilon$. It will not in general be true that a collection of quantum devices will have nearly equal $\vec \varepsilon$, but as hardware improves the difference in error between devices is likely to decrease.

\begin{figure}[t]
\centering
    \includegraphics[width=0.75\textwidth]{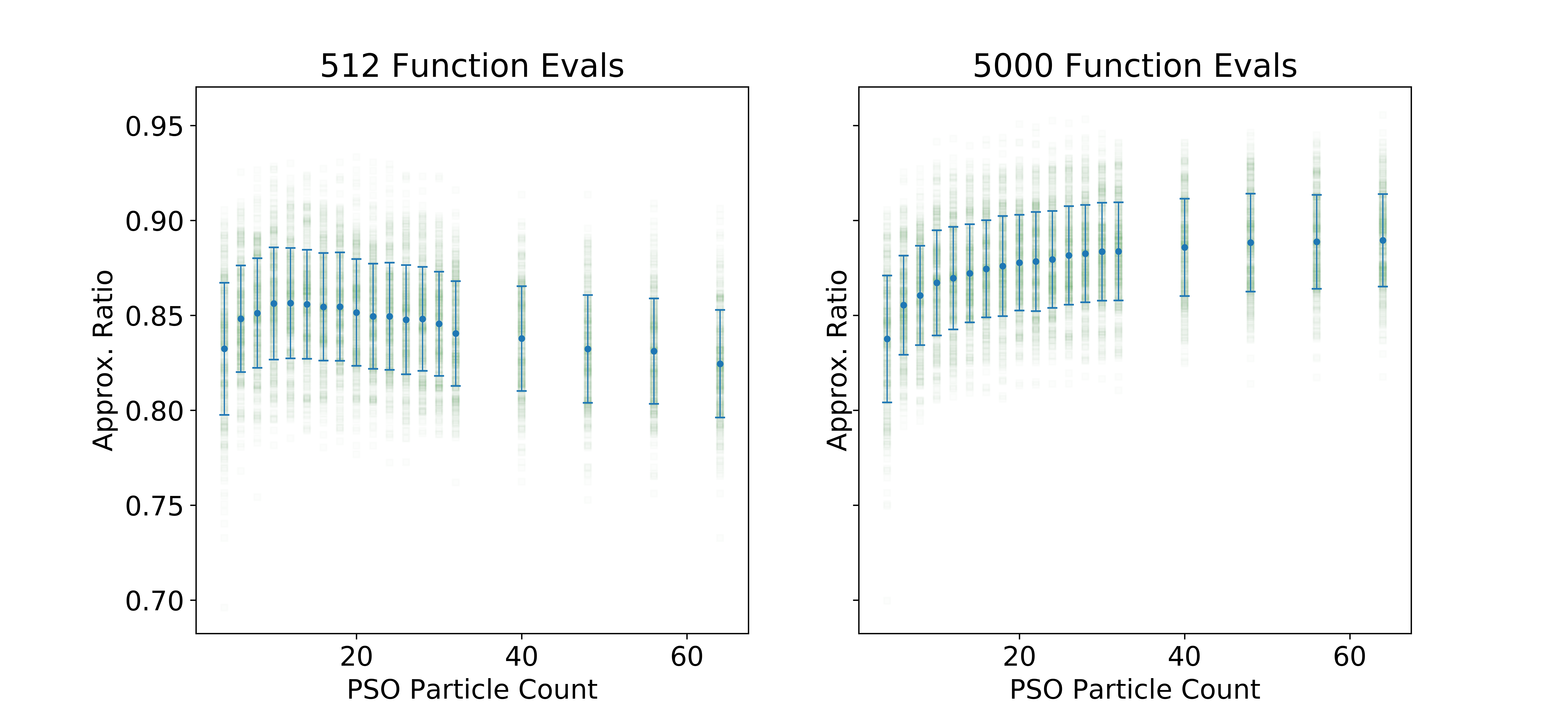}
\caption{Results of varying particle count, while running PSO for QAOA/Max-Cut on 3-regular graphs of 18 vertices (18 qubits). To compare directly between particle counts, results are shown for a fixed number of function evaluations, \textit{i.e.} a fixed number of quantum circuit emulations run on IQS. Results are averaged from 300 different graphs and random initial conditions. Error bars show standard deviation of the distribution (not uncertainty in the mean). If one is limited in the total number of functional evaluations one has the bandwidth to perform, then the utility of adding additional particles might be non-monotonic. Notably, standard deviations show substantial overlap between different particle counts. }
\label{fig:particle-count}
\end{figure}

As the number of total function evaluations $\neval$ increases, \figref{fig:swarm-dynamic} shows the improvement in the Max-Cut approximation ratio, equal to the quotient of $ \bra{0} U^\dag_{circ}$ $H_{MaxCut}$ $U_{circ} \ket{0}$ and the exact MaxCut solution $E_{exact}$. 
$H_{MaxCut}$ denotes the quantum observable that counts the number of cuts for a given graph bipartition and we are therefore interested in finding its largest eigenvalue. Results for several particle counts between 4 and 64 are shown, with the horizontal axis giving the number of function evaluations. Note that the number of function evaluations per PSO step is equal to the number of PSO particles, meaning that having more particles results in fewer swarm steps for the same number of total function evaluations.

The results show mean approximation values at each step, averaged over 300 random 3-regular graphs of 18 vertices (qubits), each with randomly selected initial conditions for the swarm. A QAOA depth \cite{Farhi14_qaoa_orig} of $p=4$ was used for all circuits, leading to $2p=8$ parameters and hence an 8-dimensional position vector. The standard deviations (from the 300 random graph instances) are omitted in \figref{fig:swarm-dynamic}, though they are appreciable (often higher than 0.05) and usually larger than the difference between means. Note that we are referring to the standard deviation of the distribution of $\mathcal L(\vec \theta) =$ $ \bra{0} U^\dag_{circ}$ $H_{MaxCut}$ $U_{circ} \ket{0}/E_{exact}$ over many graphs instances, \textit{not} the uncertainty in the mean. For the initial step, the reported values for a given $\neval$ accurately reflect the number of random positions chosen at that point. For example, for 64 particles, the value plotted at $\neval=8$ reflects the best value for the first 8 randomly chosen positions.

Though the best strategy is to choose a number of particles with the best mean behavior for a given $\neval$, the large overlaps between the standard deviations suggest that a clearly superior particle count choice would appear only after many problem instances.
%
For lower numbers of function evaluations, lower particle counts perform slightly better, because fewer function evaluations are spent on the first step of choosing many random positions. For example, at $10^2$ function evaluations, using fewer particles is always a slightly strategy. This is because there are fewer function evaluations per time step, allowing for faster convergence in the short-term. The trend is reversed well before $10^4$ function evaluations, because more space is explored by the higher particle counts. In between these extremes, one can find a $\neval$ for which an arbitrary particle count in this range performs best.

For two snapshots taken at 512, and 5000 total function evaluations, \figref{fig:particle-count} shows the mean as well as the standard deviations of the distribution. Both \figref{fig:swarm-dynamic} and \ref{fig:particle-count} show that, if one is limited by the total number of function evaluations, the optimal number of particles is not necessarily the largest number. The reason for this is, though more particles allow for exploring more of the parameter space, this comes at the cost of needing to calculate more function evaluations per swarm step. However, as $\neval$ increases, more particles ought to strictly produce better approximations to the solution, matching the observed trend.
\figref{fig:particle-count} shows more clearly that the optimal particle count depends on $\neval$. For instance, if one may only perform $\neval\sim$500 evaluations, $\sim$10-14 particles are best, but the optimal number of particles increases as the number of allowed function evaluations increases. Stated differently, the fewer total function evaluations are available, the less utility is gained from adding more particles. This relationship between total allowed function evaluations and particle count is problem-dependent, but it highlights the usefulness of classical software results when running real quantum algorithms. If one can determine optimal hyperparamters (such as particle count) for a quantum problem using classical software, it may inform the choice of hyperparameters when these problems are scaled up on real quantum hardware.



\begin{figure}[t]
\centering
    \includegraphics[width=0.63
    \textwidth]{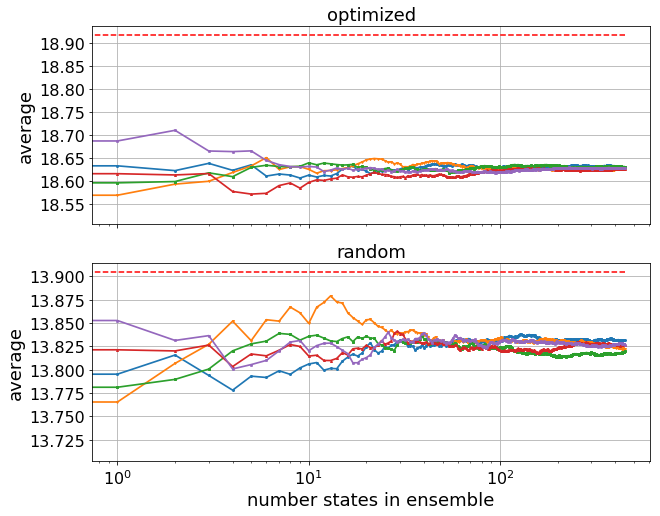}
\caption{Variational algorithms are based on the iterative improvement of quantum circuits based on the estimation of the expectation value of a specific observable. In presence of noise, the quantum state differs from the result of ideal simulations and so does the expectation value. Here we show the convergence of an ensemble of stochastic simulations to the result including noise effects. The circuit corresponds to a QAOA instance of Max-Cut on 3-regular graphs ($n=16$ qubits and 4 QAOA steps). We compiled the circuit for a device with all-to-all connectivity via a simple greedy scheduler. The data are taken for the decoherence timescale $T_1 = 500\,T_g$ and $T_2=T_1/2$, where $T_g$ is a typical gate duration. \textbf{Upper panel:} Convergence of $\bra{0} U^\dag_{circ} H_{MaxCut} U_{circ} \ket{0}$ to its noisy value by means of incoherent average over an increasing number of states in the ensemble. Each ensemble state is obtained by simulating the circuit with the addition of the noise gates according to the schedule. The circuit's parameters have been optimized with the PSO method. The different lines show the same averaging procedure but with different streams of random numbers and suggest that convergence requires hundreds of states in the ensemble, for the parameters used here. For reference, the red dashed line indicates the expectation value $\mathcal L$ for noiseless simulations. \textbf{Lower panel:} As above but with circuit's parameters initialized randomly from a uniform distribution.} 
\label{fig:noise}
\end{figure}

\subsection*{Convergence of noisy simulations}

QAOA is one of the leading candidates to achieve quantum advantage on noisy near-term quantum devices. To evaluate its performance in realistic experiments, it is fundamental to include the effect of noise and decoherence into its protocol. There are two distinct effects of noise that combine in QAOA protocols: the first one is that the expected state $U_{circ} \ket{0}$ is not achieved in practice but a hopefully related mixed state is obtained. The estimate of $\mathcal L$ on this state may differ from that of the noiseless case. Second, the imprecise value of $\mathcal L$ is transmitted to the classical optimization loop and affects the next choice of the circuit's parameters.
\figref{fig:noise} illustrates the utility of simulating multiple states in parallel to speed up noisy simulations. The QAOA/MaxCut simulations were performed on 3-regular graphs of 16 vertices, with a QAOA depth of $p=4$. We used $T_1=500T_g$ and $T_2=T_1/2$, where $T_g$ is the gate time and $T_1$ and $T_2$ are respectively the time constants related to relaxation and dephasing. In this example, an optimized set of parameters reached a converged value more quickly than a randomly selected set of parameters.


%% file: section_appendix.tex
\section{Installation}
\label{app:sec:installation}

Intel Quantum Simulator builds as a static library and it can be used via an API, which corresponds to the quantum gate operations. The following packages
are required to be present to the system, before installing the library:
\begin{itemize}
  \item CMake tools version 3.15+
  \item MPICH3 for distributed communication
  \item optional: Intel Math Kernel Libary (MKL) for distributed random number generation
  \item optional: PyBind11 (installed via conda, not pip) required by the Python binding of IQS
\end{itemize}

The code is hosted as open-source project to the public GitHub repository and it can be cloned via:
\begin{lstlisting}[language=tex,basicstyle=\small,numbers=none]
  git clone https://github.com/iqusoft/intel-qs
  cd intel-qs
\end{lstlisting}

The preferred installation for best performance takes advantage of Intel Parallel Studio compilers and is documented in the GitHub page of the project \cite{IQS}. Here we provide instructions how to use the standard GNU toolchain. The installation follows the out-of-source building and requires the creation of the directory build. This directory is used to collect all the files generated by the build process. The appropriate makefile is generated with CMake:
\begin{lstlisting}[language=tex,basicstyle=\small,numbers=none]
  mkdir build
  cd build
  CXX=g+ cmake -DIqsMPI=OFF -DIqsUTest=ON ..
  make
\end{lstlisting}

By default, MKL is not required when GNU compilers are used. The command above install the single-node version of IQS, while to install the distributed version one must set the option \snippet{-DIqsMPI=ON} instead. In this case, it is required at least the version 3.1 of MPICH for the build to be successful.

The result of the building process is twofold: on one hand the static C++ library of IQS is created as \snippet{build/lib/libintel_qs.a}, and on the other hand the executables of the unit test and several examples are saved in the folder \snippet{build/bin/}.

\section{Python bindings}
\label{app:sec:python}

In the last few years, the scientific community has adopted Python as a central language for numerical tools. In the field of quantum computing, several of the most popular frameworks have a Python interface \cite{Pennylane,Qiskit,Forest,Cirq,Azure,Projectq,Braket}. To facilitate the integration with those and other tools, we provide Python bindings of the IQS code for the single-node implementation.
By default, whenever MPI is disabled, the building process create a Python library containing the classes and methods of IQS. The library can be found in \snippet{build/lib/intelqs_py.cpython-36m-x86_64-linux-gnu.so} or in an equivalent file. The binding code itself uses the Pybind11 library which needs to be installed via \texttt{conda} (not simply with pip) to include the relevant information in CMake.
To disable the Python wrapper, even without MPI, set the CMake option selection to
\snippet{-DIqsPython=OFF}.

\section{Docker file}
\label{app:sec:docker}
The number of qubits that it is possible to simulate with Intel Quantum Simulator is constrained by the amount of memory available to hold the quantum state vector.

Cloud computing platforms make it possible for high-performance computing applications to be run on small temporary clusters of compute nodes that provide more memory than is possible on a single user's laptop or workstation. One simply allocates the compute nodes with the required memory sizes, configures them to talk to each other in a cluster, and then uses Kubernetes, SWARM, SLURM, BEEs, or some other container or cluster job orchestration package to allocate and use a Docker container or Singularity instance on each node.

In order to facilitate the use of Intel Quantum Simulator in a cloud computing multi-node configuration, we provide a Docker file that can be built to create an image that can be run on each compute node. This Docker build file downloads all of the required software packages including a host OS necessary to build and execute IQS software platform.

To use the multi-node capability of IQS in the cloud environment, it is necessary to compile the Dockerfile to run as a Singularity image.  This is due to restrictions on using SSH from within a native Docker image. Compiling down to a Singularity instance works around this restriction. 

The Docker container also provides a pre-built environment for compiling and building IQS if researchers do not have access to the correct OS version and software tools required by the platform.

\section{Parallel simulations of a pool of states}
\label{app:sec:multi-state}

The multi-state functionality is better described by providing a practical example. Here we want to simulate a 10-qubit system exposed to dissipation and decoherence, characterized by the $T_1$ and $T_2$ time respectively. The effect of noise is included by performing multiple simulations of the circuit with the addition of stochastic noise gates \cite{Bassi2008,Sawaya2016}.

The code below is a simplified version of the tutorial provided in the IQS repository (released under the Apache 2 license) and requires MPI. The circuit is trivial: for each qubit a rotation around the X axis is performed, by angles that were randomly chosen. One has control on the gate schedule and we assume that all gates are performed sequentially starting from qubit 0 until qubit 9. This is controlled by increasing the second argument of the noise gates corresponding to the duration of the simulated noise.

\begin{lstlisting}
#include "qureg.hpp" // IQS header file (additional header files may need to be included)

int main(int argc, char **argv)
{
  // Create the MPI environment, passing the same argument to all the processes.
  qhipster::mpi::Environment env(argc, argv);
  // One pool state per process. For accurate noise effects, they should be hundreds.
  int num_pool_states;
  MPI_Comm_size(MPI_COMM_WORLD, &num_pool_states);
  // Partition the MPI environment into groups of processes. One group per pool state.
  env.UpdateStateComm(num_pool_states);
  // Number of qubits, here 10.
  int num_qubits = 10;
  // Random number generator, provided by IQS.
  qhipster::RandomNumberGenerator<double> rng;
  rng.SetSeedStreamPtrs( 777777 );
  // Choose the angles of the circuit, randomly in [0,pi[.
  // They have the same value across all pool states.
  std::vector<double> angles(num_qubits);
  rng.UniformRandomNumbers( angles.data(), angles.size(), 0., M_PI, "pool");
  // Initialize the qubit register state to |00...0>
  QubitRegister<std::complex<double>> psi(num_qubits);
  psi.Initialize("base",0);
  // Associate the random number generator to the qubit register.
  // This is required by the stochastic noise gates.
  psi.SetRngPtr(&rng);
  // Set T_1 and T_2 timescale.
  double T_1=30. , T_2=15. ;
  psi_slow.SetNoiseTimescales(T_1, T_2);
  // -- Circuit simulation.
  // Each gate is preceded and followed by noise gates according to the schedule.
  for (int qubit=0; qubit<num_qubits; ++qubit)
  {
    double duration = double(1+qubit);
    psi_slow.ApplyNoiseGate (qubit, duration);
    psi_slow.ApplyRotationX (qubit, angles[qubit]);
    duration = double(num_qubits-qubit);
    psi_slow.ApplyNoiseGate (qubit, duration);
  }
  // Compute the probability of qubit 0 to be in |1>.
  double probability = psi.GetProbability(0);   
  // Incoherent average across the pool to get the noisy expectation.
  probability = env.IncoherentSumOverAllStatesOfPool<double> (probability);
  probability /= double(num_pool_states);
  return 0;
}
\end{lstlisting}

Notice that in line 15 we declare the (psudo-)random number generator included in the IQS software. It is advantageous for various scenarios that it can generate three kinds of random numbers:
\begin{description}
    \item[local] different for each pool rank (not used in the code above)
    \item[state] common to all ranks of the same state (automatically used by the noise gates)
    \item[pool] common to all ranks of the pool (used for the rotation angles of the circuit)
\end{description}
 